\documentstyle[aps,prb]{revtex}
\begin{document}
\title{Magnetostriction and magnetoelastic domains in 
antiferromagnets}
\author{ Helen Gomonay and V. M. Loktev}
\address{National Technical University of Ukraine ``KPI``, 37, ave 
Peremogy, 03056, Kyiv, Ukraine\\ e-mail: {\rm Helen Gomonay} $\langle$ 
{\tt malyshen@ukrpack.net}$\rangle$, or $\langle$ {\tt vloktev@bitp.kiev.ua}$\rangle$\\}
\date{\today}

\maketitle

\begin{abstract}
The problem of the observable equilibrium domain structure in pure
antiferromagnets (and other thermoelastics) is investigated with the use of
continuous elasticity theory. It is shown that completely rigid surface
produces the imaginary ``incompatibility elastic charges'' analogous to the
surface magnetic charges in ferromagnets. Corresponding long-range field is
shown to contribute into the ``stray'' energy of the sample that governs an
appearance of the domain structure, the contribution from the ``elastic
charges'' being proportional to the sample volume. Competition between the
elastic ``stray'' field that favors inhomogeneous strain distribution, and
external field that tends to homogenize the sample, provides the reversible
reconstruction of the domain structure under the action of external
magnetic field. 
\end{abstract}
\pacs{75.50.Ee; 75.60.Ch}

\twocolumn
\section{Introduction} The problem of antiferromagnetic (AFM) domains has a
very long history and was discussed for many times (see, e.g., Ref.
\onlinecite{Far,Erem,Bar,Bog}). This problem is tightly related with the
question of the domain structure (DS) origin in the substances that suffer
thermoelastic phase transitions, i.e. transitions at which the principal
order parameter is symmetrically conjugated with the shear components of
the strain tensor. The transitions of such a kind are usually accompanied
by appearance of the spontaneous strains along with the simultaneous change
of the electronic structure (martensitic phase transitions) or onset of the
vector or tensor order parameter of other nature (like magnetization,
polarization, etc.). In the case when the values of the principal order
parameter and/or corresponding strain tensor of low-symmetry (i.e.,
magnetic, ferrielectric, martensitic) phase are degenerated, the DS is
thermodynamically equilibrium and reversible under the action of the
external fields that can result in homogenization of the sample. The nature
of the equilibrium DS of ferro- and ferrimagnets with opposite direction of
magnetization inside the neighboring domains is well studied and is usually
related with the finite size of the sample and demagnetization effect
resulting from the surface magnetic charges. So, the question is ``What is
the reason of the clearly observed \cite{KLR,Ryab,janos2,amitin}
equilibrium DS in the cases when the noncompensated magnetization is equal
to zero or vanishingly small (e.g. in weak ferromagnets)?'' In the
Ref.\onlinecite{gomo} we have supposed that one of the possible mechanism
of equilibrium DS formation can be due to the surface effects, namely, the
surface energy decrease yielded by the global variation of the sample
shape. Although such a model has given qualitative and quantitative
agreement with the magnetostriction experiments \cite{KLR}, it application,
as it is evident from theoretical point of view, is restricted to the
rather small samples for which the surface effects can be clearly detected
\cite{1}.

In the present paper we make an attempt to generalize our previous results
\cite{gomo} and to show that the formation of equilibrium DS in the course
of thermoelastic phase transition can be explained by the effect of
low-deformable sample surface, and what is principal, corresponding
contribution into the sample energy being proportional not to the surface
area but to the sample volume.

\section{Model}
Let us consider a sample with the volume $V$ confined with the surface $S$.
Suppose, that above the critical temperature the sample is non-stressed and
non-strained. Below the critical temperature the crystal lattice becomes
unstable with respect to appearance of the spontaneous strains related with
the fluctuations of the order parameter (e.g., magnetizatioon,
polarization, etc). The spontaneous strains can result from the nonlinear
elastic effects, as in the case of martensites, as well as be related with
the appearance of spontaneous AFM moment, both cases allow no
demagnetization effects.

In the infinite sample the only equation that defines distribution of
spontaneous strain tensor $\hat u({\mathbf r})$ (${\mathbf r}$ is a space
coordinate) arises from the condition of the elastic free energy minimum
with respect to $\hat u$:

\begin{equation}
\frac{\partial F_{\rm el}(\hat u)}{\partial {\hat u}({\mathbf r})}=0, \label{1}
\end{equation}
\noindent
where $F_{\rm el}(\hat u)$ is the free energy density. Equation (\ref{1})
is equivalent to the condition of the absence of internal forces: ${\rm
div}{\hat \sigma}({\mathbf r})=0$, where $\hat \sigma$ is a stress tensor.
Elastic free energy density $F_{\rm el}(\hat u)$ is usually independent
upon the space derivatives of the strain tensor; so, the equilibrium strain
is homogeneous and thus trivially satisfies the compatibility conditions
\cite{Teo} that in approximation of linear elasticity takes a form:
\begin{equation}
{\rm inc}\hphantom{l}{\hat u} ({\mathbf r})\equiv -{\rm rot}{\rm
rot}\hphantom{l}{\hat u}({\mathbf r})=0. \label{2}
\end{equation}

If we take into account the surface of the sample, given parametrically by
equation ${\mathbf r}={\mathbf r}_S$, the situation changes in a crucial
way. The elastic properties of the sample surface differ from that of the
bulk, and homogeneous strain defined from (\ref{1}), inevitably brings
about the additional stresses in the sample. In formal, the surface can be
accounted for through the surface energy term \cite{Mar}:
\begin{equation}
F_{\rm surf}=\int dS \left[\alpha_S+{\hat \beta}_S{\hat u}({\mathbf
r}_S)\right], \label{3}
\end{equation}
\noindent
where $\alpha_S$ is a surface tension coefficient, ${\hat \beta}_S$ is the
tensor of the surface elastic modula, and ${\hat u}({\mathbf r}_S)$ is a
surface strain tensor. It should be emphasized that the first term in
(\ref{3}) is responsible for the set up of the equilibrium shape of the
sample (it establishes the minimum surface area at given volume), the
second term can be important when considering thermoelastic phase
transition.

In what follows we will suppose that the shape of the sample is given
technologically and does not relax during the time of experiment, and so,
will disregard an influence of the surface tension (\ref{3}). In other
words, we consider the energy barrier for shape variation in the strained
sample as high enough, so, that corresponding relaxation time is much
greater than the inverse energies characteristic to the phase transition.
This is the case for AFM transitions at low temperature and also for
martensitic phase transitions that usually proceed with the rate close to
sound velocity.

In the vicinity of the surface the spontaneous strains that occur during
the thermoelastic phase transition can be represented as a sum of three
summands:
\begin{eqnarray}
{\hat u}({\mathbf r})&=&\frac{1}{3}\hat {\mathbf 1}{\rm Tr}{\hat u}+{\hat
u}_{\rm shear} \nonumber\\ &+&\left\{({\mathbf n}_S{\hat u})\otimes
{\mathbf n}_S+{\mathbf n}_S\otimes ({\hat u}{\mathbf n}_S)-{\mathbf
n}_S\otimes {\mathbf n}_S({\mathbf n}_S{\hat u}{\mathbf
n}_S)\right\},\label{11}
\end{eqnarray}
where $\hat {\mathbf 1}$ is a unit matrix, ${\mathbf n}_S$ is a surface
normal. The first term in (\ref{11}) is an isotropic volume striction, the
last three terms in brackets form the component that produces the shift
along the surface normal ${\mathbf n}_S$, and the second, traceless, term
corresponds to the shear strains (with respect to ${\mathbf n}_S$). The
first component is significant at the magnetic phase transitions and can be
neglected during the martensitic ones, but at any rate it does not
contribute into the symmetry change of the crystal and hence, as well as
the last term, could not be removed by onset of DS in the sample. So, in
what follows, we consider only the shear part of the strain tensor \cite{4}
${\hat u}_{\rm shear}$ and define the strain state with respect to
isomorphous magnetostriction which contribution can be taken into account
trivially.

If one assumes that the sample surface is not deformed during the phase
transition, i.e., ${\hat \beta}_S=0$, we should impose the standard
boundary conditions for zero external stress
\begin{equation}
\left\{\frac{\partial F_{\rm el}({\hat u})}{\partial {\hat
u}}\right\}_{{\mathbf r}\in S} {\mathbf n}_S({\mathbf r}_S)=0,
\label{4}
\end{equation}
and additional conditions of the absence of surface strains, i.e.
\begin{equation}
{\mathbf n}_S({\mathbf r_S})\times \hat u({\mathbf r_S}) \times {\mathbf
n}_S({\mathbf r_S})=0,
\label{5}
\end{equation}
that can be easily deduced from the definition of the tangent component of
the strain tensor. In (\ref{4}), (\ref{5}) ${\mathbf n}_S({\mathbf r_S})$
is the surface normal at a given point.

Thus, the problem of equilibrium strain distribution in the definite-shaped
sample below the phase transition temperature should be reformulated as
follows: equations (\ref{1}),(\ref{2}) should be satisfied inside the
sample volume for ${\mathbf r}\in V \setminus S$, and equations (\ref{3}),
(\ref{4}) are fulfilled at the sample surface, ${\mathbf r}={\mathbf
r}_S\in S$.

It is obvious that for the symmetry-governed transition, onset of
homogeneous strain inevitably breaks the condition (\ref{5}). The only
exceptions being the cases of a rather thin disk- or plate-shaped sample
with the plate surface coinciding with the homogeneous shear plane
\cite{5,6}.

\section{``Stray'' energy}
The problem of the equilibrium DS in AFM can be formally solved in analogy
with the well known problem of the domain distribution in ferromagnets;
namely, we assume that equilibrium (and in general, inhomogeneous) shear
strain of the sample that complies with the condition (\ref{5}) consists of
two parts:
\begin{equation}
{\hat u}_{\rm shear}({\mathbf r})={\hat u}_{\rm ms}({\mathbf r})+{\hat
u}_{\rm ch}({\mathbf r}), \label{6}
\end{equation}
\noindent
where the first term, ${\hat u}_{\rm ms}({\mathbf r})$, corresponds to the
eigen spontaneous magnetostriction-induced strain defined from the free
energy minimum (\ref{1}), and the second term, ${\hat u}_{\rm ch}({\mathbf
r})$ is an additional strain field produced by the ``incompatibility
elastic charges'' localized at the crystal surface with the density:
\begin{equation}
{\hat e}_{\rm elas}({\mathbf r)=-{\mathbf 
n}_S\times {\hat u}_{\rm ms}({\mathbf r_S}) \times {\mathbf n}_S\delta
\prime [{\mathbf n}_S({\mathbf r}-{\mathbf r}_S})],
\label{7}
\end{equation}
\noindent
where $\delta$ is the Dirac's delta-function, prime means derivative vs
argument. Function $\hat u_{\rm ch}({\mathbf r})$ can be found explicitly
from equation
\begin{equation}
{\rm inc} \hphantom{I}\hat u_{\rm ch}({\mathbf r})=\hat e_{\rm elas}
({\mathbf r})
\label{A4}
\end{equation}
as follows (for the details see Ref.\onlinecite{Teo})
\begin{eqnarray}
\hat u_{\rm ch}({\mathbf 
r})&=&\frac{1}{4\pi}\int_{V}d{\mathbf r}_1\frac{\hat e_{\rm elas}({\mathbf
r}_1)-\hat 1 {\rm Tr} {\hat e_{\rm elas}} ({\mathbf r}_1)}{\mid {\mathbf
r}-{\mathbf r}_1\mid}
\nonumber\\
&=&\frac{1}{4\pi}\int_S dS \frac{({\mathbf n},{\mathbf r}-{\mathbf
r}_S)}{\mid {\mathbf r}-{\mathbf r}_S \mid^3}{\hat {\mathbf U}} ({\mathbf
r}_S)\nonumber\\ &=&\frac{1}{4\pi}\int d\Omega_{\mathbf r}\hat {\mathbf
U}({\mathbf r}_S), \label{A5}
\end{eqnarray}
\noindent
where 
\begin{eqnarray}
{\hat U_S}({\mathbf r}) \equiv \hat u_{\rm ms}({\mathbf r})+{\mathbf
n}_S\otimes {\mathbf n}_S {\rm Tr} \hat u_{\rm ms}({\mathbf r})\nonumber \\
-{\mathbf n}_S\otimes (\hat u_{\rm ms}({\mathbf r}) {\mathbf n}_S)-( \hat
u_{\rm ms}({\mathbf r}){\mathbf n}_S) \otimes  {\mathbf n}_S.\nonumber
\end{eqnarray} 
\noindent
and $d\Omega_{\mathbf r}$ is an increment of solid angle at which the
surface point ${\mathbf r}_S$ is seen from the given point ${\mathbf r}$.

Strain $\hat u_{\rm ch}({\mathbf r})$ produces the ``twinning`` stress
\begin{equation}
{\hat \sigma}({\mathbf r}) = \hat{\hat c} \hat u_{\rm ch}({\mathbf r}),
\label{A6}
\end{equation}
\noindent
completely analogous to demagnetization field in ferromagnets. In (\ref{A6}) 
$\hat {\hat c}$ is the tensor of the elastic modula. Equilibrium strain
distribution conditioned by nondeformed surface can thus be found from
equation:
\begin{equation}
\frac{\partial F_{\rm el}({\mathbf r})}{\partial {\hat u}(\mathbf r)}=
\hat{\hat c} \hat u_{\rm ch}({\mathbf r}),
\label{8}
\end{equation} 
that combines (\ref{A6}) and (\ref{1}).
It is easy to see that (\ref{8}) can be also deduced from variation of free
energy functional with respect to the strain tensor components that satisfy
compatibility conditions (\ref{1}) inside the sample \cite{7}:
\begin{equation}
\Phi=\int_{V}d{\mathbf r}F_{\rm el}({\mathbf r})-F_{\rm stray}, \label{9}
\end{equation}
where 
\begin{equation}
F_{\rm stray}=\frac{1}{4\pi}\int_{V} d{\mathbf r} \int_S dS 
\frac{({\mathbf n},{\mathbf r}-{\mathbf r}_S)}{\mid ({\mathbf r}-{\mathbf
r}_S)\mid ^3}\hat u({\mathbf r}) \hat {\hat c} \hat U ({\mathbf r}_S)
\label{A7}
\end{equation}
is a twinning (stray) energy that was not considered in
Ref.\onlinecite{gomo}.

The above-developed approach makes it possible to solve the complicated
problem of the domain distribution in the nonlinear media under the action
of external field in terms of strain tensor which in contrast to the shift
vector is observable and symmetrically related with an order parameter.

To continue the analogy with the problem of ferromagnetic domains, we
should emphasize some properties of the twinning (stray) fields in
thermoelastics. First, as it is clearly seen from (\ref{A5}),
``incompatibility charges'' can produce homogeneous additional strain $\hat
u_{\rm ch}$ and thus, a macrostress in the case of appropriate form of the
sample. What is important, the energy contribution resulting from the
macrostress is proportional to the sample volume $V$ and thus is principal,
regardless of sample size. On the contrary to ferromagnets, where
demagnetization charges are the scalars, in the case of elasticity, the
charges (see (\ref{7})) possess the tensor characteristics, so, the only
obvious case for homogeneous twinning stress is the thin plate. This case
will be considered in details below. The other interesting feature of the
additional strain field (\ref{A5}) is that it is scaling-invariant. In
other words, additional strain distribution inside the sample depends only
upon the angle at which the surface is seen from the given point. So,
isomorphic transformation of the sample does not change the additional
strain distribution.

From the above considerations we can make rather general conclusion: in the
case of temperature-induced phase transition the macroscopic symmetry of
low- and high temperature phases is the same. If the transition is
symmetry-breaking (at the microscopic scale), i.e. the microscopic order
parameter is conjugated with nonisomorphous striction and produces the
strains that locally reduce the symmetry of the crystal lattice, then, the
macroscopic symmetry is restored due to the onset of the DS with the
differently oriented strain tensors.

\section{ Equilibirum domain structure}
Indeed, expressions (\ref{9}) and (\ref{A7}) show that the shape of the
sample essentially affects the strain distribution below the phase
transition point. The simplest way to elucidate this point is to consider
the case of thin plate for which the additional strains and twinning
stresses are homogeneous. Suppose, the plate surface is ${\mathbf
n}_S$-oriented with respect to crystal axes in high-symmetry phase, the
transverse dimensions $L$ of the plate  are much greater than its thickness
$d$. The main contribution to the stray energy (\ref{A7}) arises from the
averaged magnetostriction:
\begin{equation}
\langle {\hat u}_{\rm ms} \rangle=\frac{1}{V}\int_VdV{\hat u}_{\rm
ms}({\mathbf r}). \label{19}
\end{equation}
Substituting (\ref{19}) into (\ref{A5}) and (\ref{A6}), one obtains a
simple expression for the ``charge''-induced macrostresses:
\begin{equation}
{\hat \sigma}_{\rm macro}=\hat {\hat c}\langle \hat {\mathbf U}({\mathbf
r}_S) \rangle. \label{20}
\end{equation}
Altimately, from (\ref{20}) and (\ref{A7}) stray energy is obtained as
 follows:
\begin{eqnarray}
F_{\rm stray}&=&L^2d\left\{\langle {\hat u}_{\rm ms} \rangle \hat 
{\hat c} \langle {\hat u}_{\rm ms} \rangle 
+ {\mathbf n}_S\otimes {\mathbf n}_S{\rm Tr}\langle {\hat u}_{\rm ms} 
\rangle\right.\nonumber\\
&&\left. -{\mathbf n}_S\otimes (\langle {\hat u}_{\rm ms} \rangle {\mathbf
n}_S)-(\langle {\hat u}_{\rm ms} \rangle {\mathbf n}_S)\otimes {\mathbf
n}_S\right\}. \label{10}
\end{eqnarray}
If we restrict ourselves with the shear strains ${\hat u}_{\rm ms}={\hat
u}_{\rm shear}$, as it was assumed above (see (\ref{11}) and (\ref{6})),
then the stray energy takes a form
\begin{equation}
F_{\rm stray}=L^2d\langle {\hat u}_{\rm ms}\rangle {\hat{\hat c}}\langle
{\hat u}_{\rm ms} \rangle \geq 0,
\label{12}
\end{equation}
that explicitly reveals it positivity for any choice of shear strain. In
analogy with the case of ferromagnets, in the absence of external field the
stray energy $F_{\rm stray}$ can only be reduced by zeroing of the average
strains. Expression (\ref{12}) coincides  with the expression (2) of
Ref.\onlinecite{gomo} written previously from the phenomenologic
considerations only.

The value of $F_{\rm stray}$ (\ref{12}) obviously depends upon the shape of
the sample, in our case, from the orientation of plate surface with respect
to crystal axes. If the plate normal ${\mathbf n}_S$ is directed along the
principal symmetry axis $C_n$ of the sample ($n$=3, 4, or 6) then, the
macroscopic symmetry of the sample must be restored below the phase
transition point and all the types of the domains should be equally
represented.

The above calculations convincingly show that appearance of the homogenous
deviatoric (shear) strains in the finite-size sample can give rise to a
considerable energy increase and thus are non-advantageous. There is a
close analogy between the appearance of the long-range elastic fields in
thermoelastics and dipole fields in ferromagnets (a similarity between the
equations for electro-magnetic fields in substances and equations of
elasticity theory was long ago noticed by R. de Wit in
Ref.\onlinecite{Wit}). The case of the weak ferromagnets needs the special
treatment because the DS in corresponding compounds can be formed due to
competition of demagnetization and twinning factors. This problem is out of
scope of this paper.

The general expression (\ref{12}) can also be used for the description of
the martensitic phase transitions with the only corrections concerning
adjustable habitus (zero-shift) plane instead of fixed sample surface.

Another important question that arises while analyzing the expression
(\ref{10}) is whether onset of twin (domain) structure is indeed
thermodynamically advantageous. The matter is that the inhomogeneous strain
distribution gives rise to an increase of the full free energy due to the
contribution of the domain walls. In order to show that the energy gain
(\ref{12}) because of the twinning is greater than the loss from the domain
walls contribution, let us follow the method applied for the ferromagnets
and consider the DS consisting of two alternating types of the domains,
characterized with the strain tensors ${\hat u}_1$ and ${\hat u}_2$, the
period of the structure $d_{\rm DS}$ is much less then  the plate
thickness, $d_{\rm DS} \ll d$. We suppose that the domains are coherently
conjugated with each other and go out to the surface with the same strains
that in the bulk.

Standard calculations based on the formulae (\ref{9}) and (\ref{A7}) give
rise to the following contribution into free energy from the ``elastic
charges'' \cite{8}
\begin{equation}
F_{\rm ch}=L^2d_{\rm DS} \xi_1 \xi_2({\hat u}_1-{\hat u}_2)\hat {\hat
c}({\hat u}_1-{\hat u}_2) \cos \vartheta, \label{13}
\end{equation}
where $\vartheta$ is the angle between the plate surface and domain (twin)
interface, $\xi_1(=1-\xi_2)$ is the volume fraction of the domain of the
first type.

Domain walls (interfaces) also contribute into the free energy,
corresponding expression is:
\begin{equation}
F_{\rm DW}=L^2\sigma_{\rm DW}\frac{d}{d_{\rm DS}},
\label{14}
\end{equation}
where $\sigma_{\rm DW}$ is the surface energy of a single domain wall.

Comparison of equations (\ref{10}), (\ref{12}) and (\ref{13}) shows that
the energy of the domain walls compete with the contribution from the
``elastic charges'' and thus defines an optimal DS period:
\begin{eqnarray}
d_{\rm DS}^{\rm opt}=\frac{\sqrt{\sigma_{\rm DW} d}}{\sqrt{ \xi_1
\xi_2({\hat u}_1-{\hat u}_2)\hat {\hat c}({\hat u}_1-{\hat 
u}_2) \cos \vartheta }}\nonumber\\
\simeq \sqrt{\frac{\sigma_{\rm DW} d}{F_{\rm el}}}. \label{15}
\end{eqnarray}

So, contribution from the inhomogeneous part of strains and short-range
periodic ``charge'' distribution is proportional to the ratio $d_{\rm DS}/d
\ll 1$ and is much less than the stray energy (\ref{10}) that is
proportional to the sample volume. This means that onset of the DS in
thermoelastics crucially diminishes the free energy from the ``elastic
charges'' localized on the sample surface, the energy increase related with
the inhomogeneous strain distribution is much smaller and cannot compensate
the energy related with homogenization of the domain structure for rather
large samples \cite{9}.

\section{Concrete examples}
To elucidate the consequences of the stray energy, let us consider a case
of plate-shaped easy-plane pure AFM with the plane normal directed along
the principal crystal axis. A good example of such an AFM is given by
CoCl$_2$ (symmetry group is $D_{3d}$) and low-doped YBa$_2$Cu$_3$O$_{6+x}$
at $x \le 0.3$ (symmetry group is $D_{4h}$) single crystals. Corresponding
contribution into stray energy (\ref{10}) is:
\begin{eqnarray}
F_{\rm stray}=L^2d\left\{\frac{1}{2} c_{11}\left[\langle
u_{xx}\rangle^2+\langle u_{yy}\rangle^2\right] \right.
\nonumber\\ 
\left. +c_{12}\langle u_{xx}\rangle \langle u_{yy}
\rangle+2c_{66}\langle u_{xy}\rangle^2 \right\},
\label{A8}
\end{eqnarray}
where $z$-axis is supposed to be directed along the plate normal.
Full free energy with the account of the magnetic subsystem in the external
magnetic field ${\mathbf H}$ (neglecting the small demagnetization effects)
could be written as follows
\begin{eqnarray}
F=L^2d\sum_k \left\{\xi_k \lbrack E_{\rm an}+\frac{M_0}{2H_E}({\bf H}{\bf
l}_k)^2\right.\nonumber\\ \left.+ \hat u_k \hat{\hat \lambda}_{\rm me} {\bf
l}_k\otimes {\bf l}_k+\frac{1}{2} \hat u_k \hat {\hat c} \hat u_k
\rbrack\right\}+F_{\rm stray},\label{16}
\end{eqnarray}
where $E_{\rm an}$ is the magnetic anisotropy energy, $M_0$ is saturation
magnetization, $2H_E$ is the value of spin-flip field of exchange nature,
${\bf l}_k$ and $\hat u_k $ are the vector of AFM and spontaneous
magnetostriction, respectively, in the $k$-th domain ($k$=1, 2, 3 for
CoCl$_2$ and $k$=1, 2 for YBa$_2$Cu$_3$O$_{6+x}$), $\xi_k$ is the volume
fraction of the $k$-th domain, $\hat {\hat \lambda}_{\rm me}$ is the 4-rank
tensors of magnetoelastic coefficient, the strain tensor is averaged over
the sample volume as follows: $<\hat u>=\sum_k \xi_k \hat u_k$.

Minimization of the energy (\ref{16}) with respect to components of AFM
vectors, strain tensor and also to domain fraction $\xi_k$ proves
\cite{gomo} that due to redistribution of the domains, the effective
magnetic field inside sample is zero until the external field attains the
value
\begin{equation}
\mid {\mathbf H}\mid =H_{\rm MD}\equiv \lambda_{\rm eff}
M_0\sqrt{\frac{H_EM_0 }{c_{\rm eff}}}, \label{17}
\end{equation}
where $\lambda_{\rm eff}$ and $c_{\rm eff}$ are certain combinations of
magnetoelastic and elastic constants that should be calculated with the
account of concrete symmetry of the crystal. The value $H_{\rm MD}$ can be
associated with the field of monodomenization of the sample \cite{gomo}.

Below the filed of monodomenization the averaged shear strain
\begin{equation}
\langle u_{\rm shear}\rangle =\frac{\lambda_{\rm eff} M_0^2}{c_{\rm
eff}}\left\{\frac{H}{H_{\rm MD}}\right\}^2=u_0\left\{\frac{H}{H_{\rm
MD}}\right\}^2.
\label{18}
\end{equation}
follows the quadratic field dependence normalized to the factor of
monodomenization field value. In (\ref{18}) $u_0$ describes the absolute
value of the local spontaneous shear. Thus, the low-field dependence of any
macroscopic property which depends upon the relative fraction of the
elastic domains should follow the law of the ``corresponding states''
\cite{10}, i.e., should coincide for different substances in the reduced
coordinates $H/H_{\rm MD}$ for the case $H\le H_{\rm MD}$.

\section{Conclusions}
The main results of the paper can be formulated as follows:
\begin{enumerate}
\item The equilibrium DS that arises in the course of thermoelastic phase
transition results from the finite-size effects closely related with the
properties of sample surface. The absolutely rigid surface produces the
imaginary ``incompatibility elastic charges'' which long-range field
contributes into the energy of the sample, corresponding contribution being
proportional to the sample volume and thus provides the twinning effect.
\item Contribution from inhomogeneous distribution of the strain below the
phase transition point is proportional to the DS period and in the case of
large samples is vanishingly small.
\item Elastic stray energy is the reason of the onset of equilibrium DS in
pure AFM. The domain distribution can be reversibly regulated by the
external magnetic field.
\item Low-field dependence of macroscopic parameters should be the same for
different samples if compared in the reduced coordinates $H/H_{\rm MD}$
(the law of the ``corresponding states'').
\end{enumerate}
\acknowledgements
The authors are grateful to Profs. M.A.Ivanov, V.I.Marchenko and
S.M.Ryabchenko for numerous discussions on the problem of mechanism of the
DS formation in AFM.


\begin{thebibliography}{99}
\bibitem{1}
Strictly speaking, the surface effects as treated in Ref.\onlinecite{gomo},
can give rise to equilibrum DS only for special geometry of the sample,
namely, for thin plate or disk with the normal taken along the principal
crystal symmetry axis.
\bibitem{4}
Another point is that the isomorphous striction and shear strain are
transformed according to different indecomposable representations of the
crystal symmetry group. Separated consideration of both constituents means
that we neglect in free energy of the crystal anharmonic terms which are
usually small.
\bibitem{5}
This case is realized, for example, in the pure cobalt where fcc-hcp phase
transition takes place. The both phases can be transformed to each other by
the shift of closely-packed planes, so no incompatibility occurs on the
interface between the fcc and hcp phases.
\bibitem{6}
Another case of the thin plate with the normal in the direction of
spontaneous elongation (contraction) seems to be unreal for the traceless
part of strain tensor. Really, elongation (contraction) in plane normal
direction correspond to contraction (elongation) in perpendicular direction
lying in the plate plane and thus breaking the condition (\ref{5}).
\bibitem{7}
Equations (\ref{8}) in the divergent form can also be obtained from the
functional (\ref{9}), (\ref{A7}) by varying it with respect to the shift
vector components. The compatibility conditions in this case are obviously
excessive.
\bibitem{8}
An analogous expression was deduced in Ref.\onlinecite{Roi} on the basis of
the empirical Saint-Venan principle.
\bibitem{9}
If the sample thickness is of the order of optimal DS period, the
preferable configuration is homogeneous.
\bibitem{10}
A kind of such conformity was found out in Ref.\onlinecite{Ryab} for the
high-filed dependence of magnetostriction where the sample can be treated
as monodomain, the characteristic field being the spin-flip field value and
corresponding dimension-less coordinates were $H/(2H_E)$.
\end{thebibliography}
\end{document}